# All-antiferromagnetic electrically controlled memory on silicon featuring large tunneling magnetoresistance

*Jiacheng Shi[1], Victor Lopez-Dominguez[1,2]\*, Sevdenur Arpaci[1,3], Vinod K. Sangwan[4], Farzad Mahfouzi[5], Jinwoong Kim[5], Jordan G. Athas[1], Mohammad Hamdi[1], Can Aygen[1], Charudatta Phatak[6], Mario Carpentieri[7], Jidong S. Jiang[6], Matthew A. Grayson[1,3], Nicholas Kioussis[5]\*, Giovanni Finocchio[8]\*, Mark C. Hersam[1,3,4,9], Pedram Khalili Amiri[1,3]\**

**Affiliations:**

[1] Department of Electrical and Computer Engineering, Northwestern University; Evanston, Illinois 60208, United States of America.

[2] Institute of Advanced Materials (INAM), Universitat Jaume I; Castellón, 12006, Spain.

[3] Applied Physics Program, Northwestern University; Evanston, Illinois 60208, United States of America.

[4] Department of Materials Science and Engineering, Northwestern University; Evanston, Illinois 60208, United States of America.

[5] Department of Physics and Astronomy, California State University Northridge; Northridge, California 91330, United States of America.

[6] Materials Science Division, Argonne National Laboratory; Lemont, Illinois 60439, United States of America.

[7] Department of Electrical and Information Engineering, Politecnico di Bari; Bari, Italy.

[8] Department of Mathematical and Computer Sciences, Physical Sciences and Earth Sciences, University of Messina; Messina 98166, Italy.

[9] Department of Chemistry, Northwestern University; Evanston, Illinois 60208, United States of America.

\*Email: victor.lopez@uji.es, nick.kioussis@csun.edu, gfinocchio@unime.it, pedram@northwestern.edu

**Abstract:** Antiferromagnetic (AFM) materials are a pathway to spintronic memory and computing devices with unprecedented speed, energy efficiency, and bit density. Realizing this potential requires AFM devices with simultaneous electrical writing and reading of information, which are also compatible with established silicon-based manufacturing. Recent experiments have shown tunneling magnetoresistance (TMR) readout in epitaxial AFM tunnel junctions. However, these TMR structures were not grown using a silicon-compatible deposition process, and controlling their AFM order required external magnetic fields. Here we show three-terminal AFM tunnel junctions based on the noncollinear antiferromagnet $PtMn_3$, sputter-deposited on silicon. The devices simultaneously exhibit electrical switching using electric currents, and electrical readout by a large room-temperature TMR effect. First-principles calculations explain the TMR in terms of the momentum-resolved spin-dependent tunneling conduction in tunnel junctions with noncollinear AFM electrodes.

**Introduction**

Antiferromagnetic (AFM) materials have unique advantages such as robustness against external magnetic fields, high-frequency dynamics at picosecond time scales, and vanishing net macroscopic magnetization, attracting the interest of the semiconductor industry. These properties may enable the next generation of memory and computing devices with unprecedented speed, energy efficiency, and bit density, as well as resonant electrically tunable terahertz detectors and emitters [1-10].

Recent studies have established the possibility of using electric currents, by means of spin-orbit torque (SOT), to manipulate magnetic order in AFM thin films and heterostructures [5, 11-23]. However, the practical realization of AFM memory devices requires mechanisms for *both manipulation and detection* of AFM order by electrical means. Until recently, the electrical detection of the magnetic state in AFM structures exclusively relied on the anisotropic magnetoresistance (AMR) and spin Hall magnetoresistance (SMR) effects, which provided relative resistance variations ($\Delta R/R$) that were too small for memory applications [24, 25].

One approach to solving this readout issue is to utilize tunneling effects, which in the case of ferromagnetic (FM) devices, have been previously shown to provide substantially larger $\Delta R/R$ values [26, 27]. In addition, this type of readout can, in principle, be combined with SOT in a three-terminal device structure, providing separate electrical paths for writing and reading of information. A recent work demonstrated that an AFM state, switched by SOT, can be imprinted on an adjacent ferromagnetic layer for readout by TMR [23]. The presence of a ferromagnetic tunnel junction within the device, however, eliminates some of the potential advantages of an AFM memory listed above. Realizing large TMR effects in all-AFM tunnel junctions, however, is a significantly greater challenge compared to the ferromagnetic case, due to the absence of macroscopic magnetization in antiferromagnets.

Two general approaches have been proposed to address this challenge. The first involves so-called altermagnetic materials [28, 29], which have collinear moments and a staggered spin structure in both real and momentum space, giving rise to unconventional spin current generation in the presence of electric charge currents [30, 31]. The second approach, adopted here, involves noncollinear antiferromagnets of the $XMn_3$ family. These materials feature a helicity of the spin polarization in momentum space, which was recently predicted to result in sizeable dependence of the tunneling conduction on the Néel vector orientation in noncollinear AFM-based tunnel junctions, *i.e.*, the fully antiferromagnetic analogue of the TMR effect [32].

While this latter effect has been recently observed in all-AFM tunnel junctions based on $PtMn_3$ [33] and $SnMn_3$ [34], these experiments relied on fabrication processes that are difficult to scale, namely epitaxially grown single-crystal films on non-silicon substrates ($MgAl_2O_4$ and MgO, respectively). In addition, their AFM order was manipulated by using large external magnetic fields, making them impractical for electronic memory devices, which require an all-electrical read and write protocol.

Here we demonstrate the first all-antiferromagnetic tunnel junction (ATJ) devices with high TMR, grown via a scalable sputter deposition on conventional thermally oxidized silicon substrates. The three-terminal device has an all-electrical read and write protocol, whereby the magnetic state of the non-collinear antiferromagnetic $PtMn_3$ free layer is electrically written by current-induced SOT, in the absence of any external magnetic fields, and is subsequently read by using the TMR effect. The devices exhibit SOT-controlled TMR ratios as large as 110%. In this device structure, illustrated in Fig. 1, an $Al_2O_3$ tunnel barrier separates a fixed $PtMn_3$ top AFM layer from a free $PtMn_3$ bottom AFM layer, which has a predominant [111] texture due to being grown on a Pt seed layer, which also acts as the source of SOT. All layers are grown using sputter deposition at room temperature. We compare these results to AMR-based differential voltage measurements in the same device structure using a previously developed measurement protocol [17], which reveal a $10^3 \times$ enhancement of the room-temperature resistive readout signal when using the TMR readout mechanism. Additional current-induced switching measurements, performed on reference tunnel junction devices without the AFM elements, confirm the magnetic origin of the switching signal.

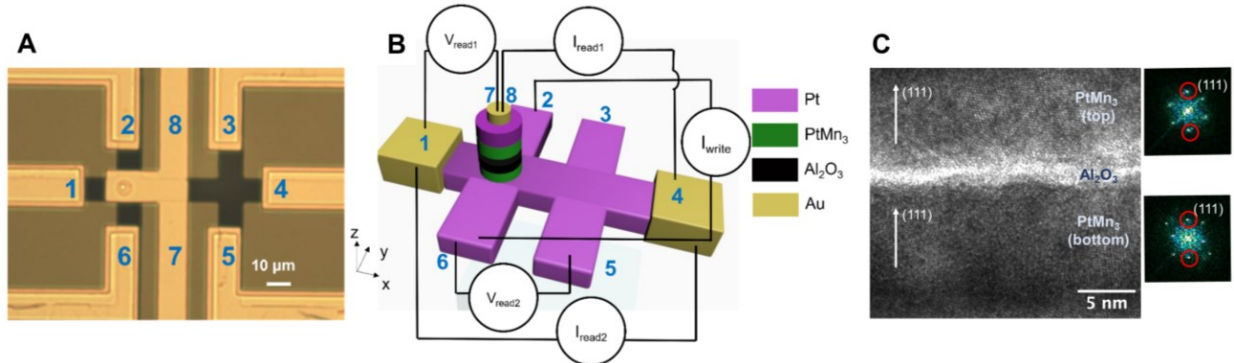

**Fig. 1. Layout and device structure of the $PtMn_3$-based three-terminal antiferromagnetic tunnel junctions. A,** Optical microscope image of the device. Note that the tunnel junction is covered by the top electrodes 7 and 8. Electrodes 1 through 6 are bottom electrodes. **B,** Schematic and measurement configuration of the three-terminal tunnel junctions. The write current is applied between electrodes 2 and 6 with opposite directions, exerting spin-orbit torque on the bottom $PtMn_3$ free AFM layer, which is in direct contact with the underlying Pt layer. Terminals 3 and 5 are used for control experiments to verify that the observed switching signals originate from the $PtMn_3$. Read currents $I_{read1}$ and $I_{read2}$ are applied to sense the average Néel vector configuration using tunneling resistance (TR) and differential voltage (DV) measurements, respectively. **C,** High-resolution transmission electron microscopy (HRTEM) image of the cross-section of the $PtMn_3/Al_2O_3/PtMn_3$ structure, showing the [111] preferred orientation of the grains in each $PtMn_3$ layer determined from the diffractograms.

## Origin of TMR in $PtMn_3/Al_2O_3/PtMn_3$ tunnel junctions

Unlike conventional ferromagnet-based tunnel junctions where TMR can be straightforwardly understood in terms of the free layer magnetization direction, the chiral nature of noncollinear (NC) antiferromagnets can also give rise to TMR, even in the absence of a net macroscopic magnetization. The necessary ingredients for TMR in spin-neutral tunnel junctions containing NC-AFM layers are (*i*) the non-relativistic momentum-dependent spin polarization resulting from the noncollinear magnetic ordering [32, 33], and (*ii*) the conservation of the tunneling electron's momentum parallel to the transport direction, $\vec{k}_{\parallel}$. In this case, the spin polarization of the tunneling current at each $\vec{k}_{\parallel}$-point contributes a finite, and overall positive, value to the total TMR.

Figure 2 shows the spin polarization of the Fermi surface in bulk PtMn$_3$. Fig. 2A shows the Fermi surfaces, where the color intensity shows the amplitude of the expectation value of the spin operator. We focus on the Fermi surface of the band which exhibits the strongest spin polarization amplitude as shown in Fig. 2B, where the color intensity illustrates the amplitude of the projection of the electronic eigenstates on the three Mn atoms in the unit cell. Here, the magnetic moments of Mn$_1$, Mn$_2$ and Mn$_3$ atoms are oriented either parallel (state 1) or anti-parallel (state 2) to the $[\bar{2}11]$, $[1\bar{2}1]$, and $[11\bar{2}]$ directions, respectively. It is, therefore, expected that the spin polarizations of the eigenstates follow the projection value of the state on each Mn atom. As an example, the large amplitude in the spin polarization, $\langle\sigma_x\rangle$ (i.e., dark blue region on the left panel in Fig. 2D) is associated with the Mn$_2$ and Mn$_3$ atoms. The same region exhibits negative half amplitude (pale red color) in $\langle\sigma_y\rangle$ and $\langle\sigma_z\rangle$ projections, consistent with the direction of a mutual net magnetic moment, $\vec{M}(\text{Mn}_2) + \vec{M}(\text{Mn}_3) = -\vec{M}(\text{Mn}_1) \parallel [2\bar{1}\bar{1}]$. The noncollinear spin configuration of the Mn atoms is then expected to yield momentum-dependent spin polarization which is shown in Figs. 2D and 2F, corresponding to the two opposite magnetic states depicted in Figs. 2C and 2E, respectively.

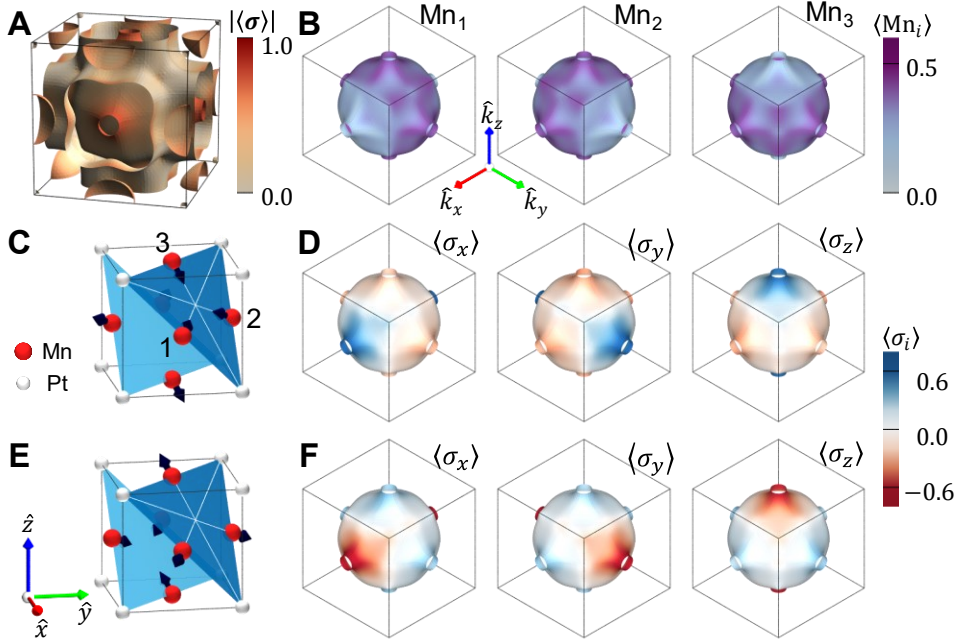

**Fig. 2. Fermi surface and spin/atom-projected density of states in PtMn$_3$. A,** Fermi surfaces in the first Brillouin zone with net spin polarization shown as color intensity. **B,** Projected orbital characters at each Mn site (Mn$_1$, Mn$_2$, Mn$_3$) of the selected band with the highest spin polarization amplitude. **C, E,** Magnetic configurations of states 1 and 2, corresponding to opposite magnetizations of all three sublattices, respectively. The black arrows of the Mn$_i$ ($i = 1,2,3$) atoms denote the magnetic moment along the $[\bar{2}11]$, $[1\bar{2}1]$, and $[11\bar{2}]$ directions (white lines). **D, F,** Projected spin texture ($\langle\sigma_x\rangle$, $\langle\sigma_y\rangle$, $\langle\sigma_z\rangle$) on the selected band of states 1 and 2, respectively.

To evaluate the expected TMR ratio emerging from this momentum-dependent spin polarization mechanism in NC-AFM tunnel junctions, we performed *ab initio* electronic structure calculations for a tunnel junction composed of PtMn$_3$ electrodes and a 1.7 nm α-Al$_2$O$_3$ barrier. The spin configuration was initialized and constrained in the [111] plane, as shown in the relaxed structure depicted in Fig. 3A. We employed the Landauer-Buttiker expression [35] to calculate the transmission across the tunnel junction for both parallel ($T_P$) and anti-parallel ($T_{AP}$) cases. Previous *ab initio* calculations [32, 33] of TMR in noncollinear AFM tunnel junctions employed as a barrier either vacuum or a monolayer of HfO$_2$. However, these choices might not accurately represent the $\vec{k}_\parallel$-resolved electron tunneling phenomenon. In contrast, the calculations reported in this work utilize an α-Al$_2$O$_3$ barrier which aligns more closely with our experimental structure. The results for the transmissions as a function of the shift of the chemical potential in PtMn$_3$ in these two magnetic configurations are depicted in Fig. 3B, as blue and green lines, respectively. The inset shows the corresponding TMR values computed from TMR = $(T_P - T_{AP})/T_{AP}$. The result shows a positive TMR up to 300%, with an average of 100% in the undoped PtMn$_3$ case. Figs. 3C and 3D show the *k*-resolved transmission in the parallel and anti-parallel cases, respectively. The results suggest that transmissions have 6-fold ($C_6$) symmetry consistent with the hexagonal symmetry of the crystal structure and are peaked near the Γ-point. In Fig. 3E, we present the *k*-resolved TMR, $\text{TMR}_{\vec{k}} = (T_P^{\vec{k}} - T_{AP}^{\vec{k}})/T_{AP}$. The result demonstrates that there is a cancellation of the transmission for *k*-points closer to the Γ-point, due to the absence of spin polarization near the Γ-point.

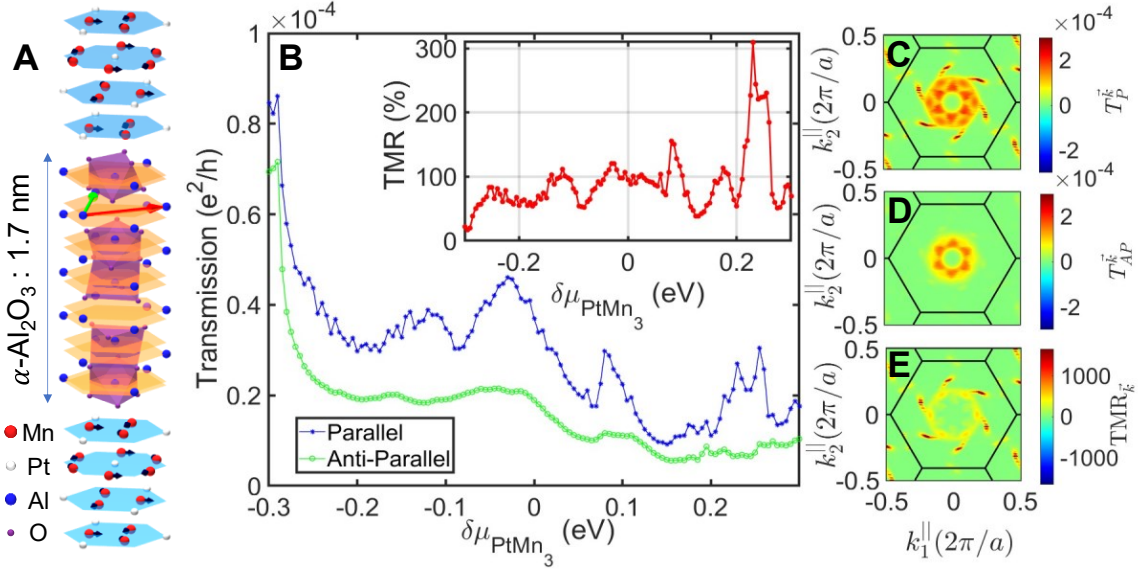

**Fig. 3. Crystal structure and *ab initio* results for TMR in PtMn$_3$[111]/Al$_2$O$_3$/PtMn$_3$[111] tunnel junctions. A,** Relaxed crystal structure of the 1.7 nm α-Al$_2$O$_3$ barrier sandwiched between two semi-infinite PtMn$_3$ leads grown along the [111] direction. The magnetic moments of the noncollinear AFM are shown as arrows in the parallel (P) configuration. In the anti-parallel (AP) configuration, the magnetic moments of the bottom PtMn$_3$ film are flipped. **B,** *Ab initio* results for the transmission through the tunnel junction for P (blue) and AP (green) configurations versus the chemical potential of PtMn$_3$. The inset (red) shows the calculated TMR as a function of the chemical potential of PtMn$_3$. **C, D,** *k*-resolved transmission at the Fermi level, $\delta\mu_{PtMn_3} = 0$, for parallel and anti-parallel configurations, respectively. **E,** *k*-resolved TMR at the Fermi level calculated by the difference of the *k*-resolved transmissions in P and AP configurations, divided by the total transmission in the AP case. The thick black lines depict the edge of the first Brillouin zone.

## Device structure and measurement configurations

Figures 1A and 1B show an optical micrograph and a schematic of the three-terminal ATJ devices. Here, the main device structure is composed of the three terminals 2, 6, and 7 (or 8), while all other terminals are used for control experiments, as outlined below. The material stack consists of a Pt(5)/PtMn$_3$(10)/Al$_2$O$_3$(2)/PtMn$_3$(10)/Pt(5) layered structure (thicknesses in brackets are expressed in nanometers), where all materials are sputter-deposited on a thermally oxidized silicon substrate, making them compatible with conventional semiconductor manufacturing processes. The devices were constructed by patterning the as-deposited ATJ films by photolithography into pillars with diameters of 6 and 8 μm on top of the bottom Pt(5) layer, which was patterned into a double-cross structure with six electrodes. The PtMn$_3$ tunnel junction pillar was placed at the center of one of the crosses (between electrodes 2 and 6, as shown in Fig. 1B). The bottom Pt layer acts as an SOT source to electrically manipulate the PtMn$_3$ AFM order, when a write current pulse of 1 ms width is applied between electrodes 2 and 6, as shown in Fig. 1B.

Figure 1C shows the high-resolution transmission electron microscopy (HRTEM) image of the ATJ stack. It can be observed that the Al$_2$O$_3$ layer is amorphous and continuous, although with relatively large roughness. The average thickness of the Al$_2$O$_3$ film is close to 2 nm. The PtMn$_3$ films are polycrystalline. The diffractograms from various grains in the PtMn$_3$ films (top and bottom) were analyzed, two of which are shown in Fig. 1C. From these diffractograms, it can be determined that the grains have a preferred [111] orientation along the growth direction. The PtMn$_3$ free layer has a non-collinear L1$_2$ phase and a Mn:Pt ratio of ~ 3.2, as revealed by X-ray diffraction (XRD) and X-ray photoelectron spectroscopy (XPS) measurements. The AFM character of the PtMn$_3$ layer was confirmed by characterizing the exchange bias in a thin Co layer adjacent to it. These structural and magnetic characterization data are provided in Supplementary Note 1.

Two types of measurements were used to read out the state of the AFM pillar in this device: (*i*) Tunneling resistance (TR) measurements, where a read current $I_{read1}$ is applied through the tunnel junction via terminals 4 and 7, while the readout voltage $V_{read1}$ is measured using terminals 1 and 8; (*ii*) Differential voltage (DV) measurement, where a read current $I_{read2}$ is applied through the underlying Pt via terminals 1 and 4, while the readout voltage $V_{read2}$ is measured using terminals 5 and 6, following a previously developed differential measurement protocol [17]. The DV measurements were used as an independent readout method, which confirms the magnetic origin of the observed current-induced switching signals in our devices. The separation of the electrical read and write paths of the device, as indicated in Fig. 1B, is beneficial for memory applications as it prevents deterioration of the tunneling barrier due to repeated write attempts, while also reducing the chance of changing the magnetic state during reading (*i.e.*, read disturbance).

## Tunneling resistance measurement results

The tunneling resistance was measured at room temperature after each electrical switching attempt, which consisted of applying ten consecutive 1 ms writing current pulses from electrode 2 to 6 and reversing the current pulse direction, from electrodes 6 to 2. Figure 4A shows the results for the TR measurement in a device with a 6 μm diameter AFM pillar. A clear current-induced switching signal is observed for a write current amplitude of 33 mA. The device exhibits a TMR ratio ~ 110%, defined as $\Delta R/R = (R_{high} - R_{low})/R_{low}$, where $R_{high}$ and $R_{low}$ are the high and low ATJ resistance levels observed in the experiment. Using the $R_{low}$ value of ~ 60 Ω, one can estimate the resistance-area product of the junction to be ~ 1.7 kΩ-μm$^2$. We emphasize that no magnetic fields were applied during this experiment. Supplementary Note 2 shows TMR switching measurements on two additional devices, which are in qualitative agreement with the results of Fig. 4.

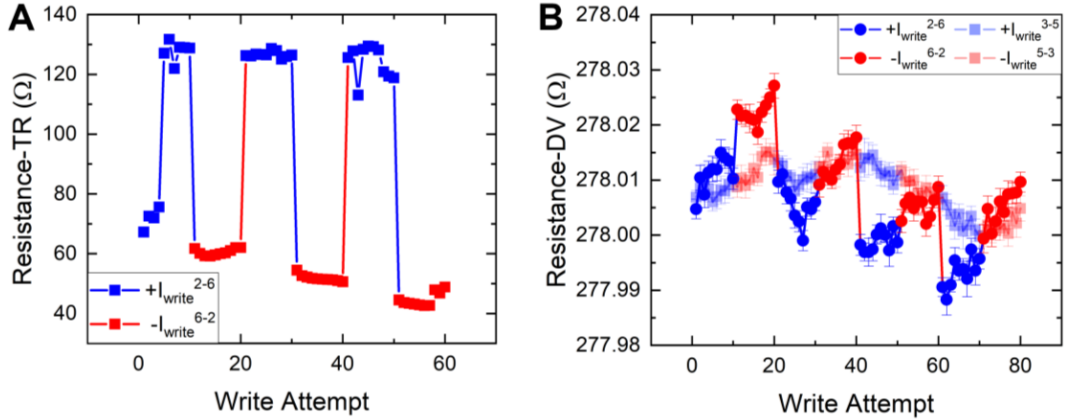

**Fig. 4. Electrical readout of the SOT-switched PtMn$_3$ state by using tunneling resistance (TR) and differential voltage (DV) measurements. A,** TR measurement results in a 6 μm diameter ATJ. Each write attempt in this case used a current amplitude of 33 mA, applied between electrodes 2 and 6. **B,** DV measurement results in the same structure (dark blue and red data points) also exhibit switching of the PtMn$_3$, but with a much smaller amplitude due to the smaller AMR effect. The background resistance of ~278 Ω is associated with the thin Pt bottom layer. Control experiments were performed in the form of DV measurements in the same device, but with write currents sent along the non-magnetic arm between electrodes 3 and 5 (light blue and red data points). As expected, no switching is observed in this case, confirming the magnetic origin of the TMR and AMR switching signals in our experiment.

**Differential voltage measurements and control experiments**

In addition to TR, DV measurements were performed where a reading current of 10 μA was applied from electrode 1 to 4, and the differential voltage was measured between electrodes 5 and 6, thus, revealing changes in the AFM Néel vector via the AMR effect. The DV measurement provides an independent test to rule out non-magnetic artifacts in the resistive switching signal [17, 24, 25]. In fact, when repeating the switching experiments using the same write current in terminals 3 and 5 of Fig. 1, *i.e.*, without the AFM element, we observed no current-induced switching in the DV measurements. This result, which is summarized in Fig. 4B, rules out the contribution of non-magnetic artifacts in the observed switching.

Note that the background resistance observed in the DV measurements of Fig. 4B is significantly larger than in the TR experiments, due to the high resistance of the long and ultrathin Pt arm of the device from electrode 1 to 4. The resistance variation in this case is estimated to be $\Delta R \sim 20$ mΩ, which is more than three orders of magnitude smaller than that measured with the TR method, clearly confirming the superiority of the tunneling measurement as the readout mechanism.

To further confirm the magnetic origin of the TMR switching signals, we also carried out current-induced switching experiments on control tunnel junction devices without any AFM element (*i.e.*, Pt/Al$_2$O$_3$/Pt devices). The results are shown in Supplementary Note 3, indicating no switching signals in the non-magnetic control experiments, as expected.

The described switching experiments were carried out on 12 ATJs made from the same tunnel junction material stack, all of which showed qualitatively similar results using write currents ranging from 32 to 36 mA. The distributions of all $\Delta R/R$ and $\Delta R$ values obtained from the TR and DV measurements are shown in Figs. 5A and 5B, respectively, indicating an average increase of the resistive readout signal by more than $10^3\times$ when using the tunnel junction readout. The $\Delta R/R$ ratio differences between the devices can be attributed to device-to-device variations induced by the fabrication process, as well as variations in the local magnetic anisotropy and domain structure, which in turn can affect the critical current required for switching of each device, as well as the micromagnetic structure of the fixed layer. The role of nonuniformities of the $PtMn_3$ magnetic anisotropy on the observed switching characteristics is supported by micromagnetic simulations, which are shown in Supplementary Note 4.

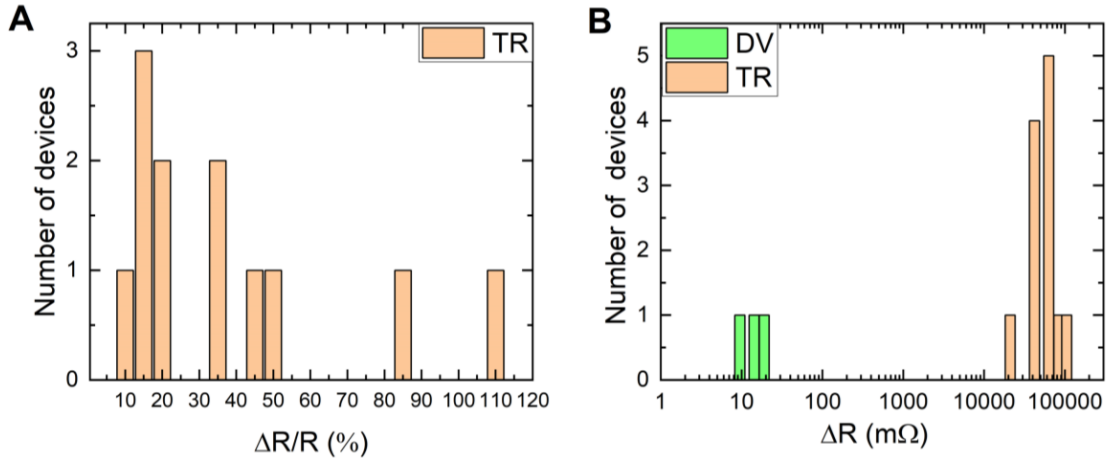

**Fig. 5. Statistics of the AFM state readout using TR and DV measurements. A,** Distribution of the $\Delta R/R$ ratio over all 12 devices measured in this work. The largest observed TMR ratio was ~ 110%. All data were measured at room temperature and in the absence of any external magnetic fields. **B,** Comparison of $\Delta R$ values from TR and DV measurements. The mean $\Delta R$ increases by more than $10^3\times$ when using the tunnel junction readout.

**Discussion and conclusions**

It is worth comparing these results to previous reports of TMR in antiferromagnet-based tunnel junctions, where manipulation of the magnetic order was performed by a magnetic field, rather than by electric currents. A related effect, tunneling anisotropic magnetoresistance (TAMR), was first experimentally observed in AFM tunnel junctions with [Co/Pt]/Ir-Mn/AlO$_x$/Pt and NiFe/Ir-Mn/MgO/Pt [36-38] material structures. In these experiments, manipulation of the AFM order was implemented using an external magnetic field, through the exchange coupling of the ferromagnetic (Co/Pt or NiFe) and antiferromagnetic (Ir-Mn) layers in the device. These structures showed TAMR only at cryogenic temperatures [36] due to the exceptionally thin AFM layers that were required to make the magnetic-field-controlled manipulation mechanism possible, thus reducing their Néel temperature below room temperature.

Two recent reports have demonstrated room-temperature TMR, controlled by external magnetic fields, in all-AFM tunnel junctions based on $PtMn_3$ [33] and $SnMn_3$ [34] noncollinear antiferromagnets. The material stacks were epitaxially grown on $MgAl_2O_4$ and MgO substrates, respectively. In all of these cases, the need for external magnetic fields to manipulate the AFM order reduced their potential as memory devices, which require an all-electrical protocol for reading and writing of information.

Another recent work demonstrated the possibility of imprinting the antiferromagnetic state of an Ir-Mn layer, which is switched by current-induced spin-orbit torque, onto a ferromagnetic CoFeB film [23]. The resulting change in the ferromagnetic order can then be read out by using conventional ferromagnetic TMR in a CoFeB/MgO/CoFeB tunnel junction. However, while providing both electrical writing and reading functions, the need for a ferromagnetic layer in these structures results in a finite magnetic dipole coupling of adjacent bits in a memory array and limits their ultimate achievable switching speed and bit density, thus partly negating the attractive features of using an AFM free layer in the device.

Thus, the three-terminal ATJs reported in the present work combine three key properties required for AFM memory device applications, which had not been realized simultaneously in previous works: Firstly, they feature room-temperature all-electrical readout of the AFM state using large TMR ratios, without the assistance of any ferromagnetic layers. Secondly, the ATJs exhibit magnetic-field-free device operation, featuring electrical control of the AFM order by current-induced SOT. Thirdly, this work demonstrates the integration of these all-AFM tunnel junction devices on a conventional silicon substrate using scalable sputter deposition.

Together, these attributes enable electrical writing and reading in an industry-relevant three-terminal device geometry with separate read and write paths. In addition to their significance as a fully functional all-AFM memory on silicon, the field-free operation of these devices also presents a significant advantage over ferromagnetic SOT memory devices, where achieving field-free operation is currently an intensely researched and challenging problem [39-47].

We expect that our demonstration of sizeable room-temperature TMR in electrically controlled ATJs on silicon will open new possibilities for a wide range of AFM spintronics experiments. Beyond memory applications, these material structures may also find applications in other devices such as sources and detectors of terahertz radiation [7, 48-50], where the output power and detection sensitivity depend on the sensitivity of electrical resistance to changes in the Néel vector.

**Acknowledgments:**

This work was supported by the National Science Foundation, Division of Electrical, Communications and Cyber Systems (ECCS-2203243, ECCS-1853879, and ECCS-1912694). This work was also supported by the National Science Foundation Materials Research Science and Engineering Center at Northwestern University (NSF DMR-1720319) and made use of its Shared Facilities at the Northwestern University Materials Research Center. This work was also supported by a research contract from Anglo American. One of the magnetic probe stations used in this research was supported by an Office of Naval Research DURIP grant (ONR N00014-19-1-2297). This work utilized the Northwestern University Micro/Nano Fabrication Facility (NUFAB), which is partially supported by the Soft and Hybrid Nanotechnology Experimental (SHyNE) Resource (NSF ECCS-1542205), the Materials Research Science and Engineering Center (NSF DMR-1720139), the State of Illinois, and Northwestern University. This work also made use of the Jerome B. Cohen X-Ray Diffraction Facility supported by the NSF MRSEC program (DMR-1720139) at the Materials Research Center of Northwestern University, and the SHyNE Resource (NSF ECCS-1542205) at Northwestern University. For part of the sample fabrication, use of the Center for Nanoscale Materials, an Office of Science user facility, was supported by the US Department of Energy, Office of Science, Office of Basic Energy Sciences, under contract no. DE-AC02-06CH11357. The work at CSUN was supported by NSF PFI-RP Grant No. 1919109, by NSF ERC-Translational Applications of Nanoscale Multiferroic Systems (TANMS) Grant No. 1160504, and by NSF-Partnership in Research and Education in Materials (PREM) Grant No. DMR-1828019. G.F. acknowledges the support from the project "SWAN-on-chip" code 101070287, funded by the European Union within the call HORIZON-CL4-2021-DIGITAL-EMERGING-01. The work of G.F. and M.C. has been also supported by the project PRIN 2020LWPKH7, funded by MUR (Ministero dell'Università e della Ricerca) within the PRIN 2020 call and Petaspin association (www.petaspin.com). V.L.-D. acknowledges the support from the Generalitat Valencia through the CIDEGENT grant CIDEXG/2022/26.

**Author contributions:**

P.K.A., V. L.-D., J.S., and S.A. designed the devices. J.S. and V.L.-D. deposited the ATJ material stacks. J.S. fabricated the devices. J.S. performed the measurements with support from V.K.S., J.G.A., S.A., M.H., C.A., J.S.J., M.A.G., and M.C.H.. F.M., J.K. and N.K. performed the *ab initio* calculations. M.C. and G.F. performed the micromagnetic simulations. C.P. performed the HRTEM analysis. J.S., V.L.-D., G.F., F.M., J.K., N.K. and P.K.A. wrote the manuscript with contributions from all the other authors. All authors discussed the results, contributed to the data analysis, and commented on the manuscript. The study was performed under the leadership of P.K.A..